\documentclass{iaus}
\usepackage{graphicx}
\usepackage{natbib}

\def\cs{$c_{\rm s}$ }
\def\solmas{$\mathrm{M_\odot}$\,}
\def\solmasp{$\mathrm{M_\odot}$}

\def\simless{\mathbin{\lower 3pt\hbox
   {$\rlap{\raise 5pt\hbox{$\char'074$}}\mathchar"7218$}}}
\def\simgreat{\mathbin{\lower 3pt\hbox
   {$\rlap{\raise 5pt\hbox{$\char'076$}}\mathchar"7218$}}}

\begin{document}

\title{Fragmentation in turbulent primordial gas}
\author[Glover et~al.]{S.~C.~O. Glover$^1$, P.~C.\ Clark$^1$,  R.~S.\ Klessen$^1$ \& V. Bromm$^2$}

\affiliation{$^1$ Institut f\"ur Theoretische Astrophysik, Zentrum f\"ur Astronomie der Universit\"at Heidelberg, Albert-Ueberle-Str.\ 2, 69120 Heidelberg, Germany \\ [\affilskip]
$^2$ Department of Astronomy, University of Texas, Austin, TX 78712 \\
}

\pubyear{2010}
\jname{Computational Star Formation}
\editors{J.~Alves, B.~Elmegreen, J.~Girart, \& V. Trimble}

\maketitle

\begin{abstract}
We report results from numerical simulations of star formation in the early universe that focus on the 
role of subsonic turbulence, and investigate whether it can induce fragmentation of the gas. We find
that dense primordial gas is highly susceptible to fragmentation, even for rms turbulent
velocity dispersions as low as 20\% of the initial sound speed. The resulting fragments cover over two
orders of magnitude in mass, ranging from $\sim 0.1$~\solmas to $\sim 40$~\solmasp. However, our
results suggest that the details of the fragmentation depend on the local properties of the turbulent velocity field and hence we expect considerable variations in the resulting stellar mass spectrum in different halos.
\keywords{stars: formation -- galaxies: formation -- cosmology: theory}
\end{abstract}

\firstsection

\section{Introduction}
\label{sec:intro}

Over the course of the last decade, work by a number of groups has lead
to the development of a widely accepted picture for the formation of the
first stars, the so-called Population III (or Pop.\ III) stars. In this
picture, the very first stars (often termed Population III.1) 
 form within small dark matter halos that have total 
masses $M \sim 10^{6} \: {\rm M_{\odot}}$, virial temperatures of around
a thousand K, and that are assembled at redshifts $z \sim 25$--30 or 
above \citep{bl04,g05,byhm09}. Gas falling into one of these small dark
matter halos is shock-heated to a temperature close to the virial temperature
of the halo, and thereafter cools via H$_{2}$ rotational and vibrational
line emission. A cold, dense core forms at the center of the dark matter halo,
with a characteristic density $n_{\rm ch} \sim 10^{4} \: {\rm cm^{-3}}$ and 
temperature $T_{\rm ch} \sim 200 \: {\rm K}$; these values are determined by the
microphysics of the H$_{2}$ molecule, and represent the point at which 
H$_{2}$ cooling becomes inefficient, and the effective equation of state of
the gas becomes stiffer than isothermal. Once the mass of this cold dense
clump exceeds the critical  Bonnor-Ebert mass for this temperature and
density, which is approximately $1000 \: {\rm M_{\odot}}$, it decouples from
the larger scale flow and undergoes runaway gravitational collapse. 
Until recently, numerical simulations that have followed this collapse up 
to very high densities \citep[e.g.][]{abn02,brlb04} have found no indication 
of further fragmentation, and so it has typically been assumed that this dense
core forms only a single star. In this picture, the final mass of this star will be
set only by the efficiency with which it can accrete gas from its natal core,
and there are good reasons to believe that for protostellar masses less than
$\sim 20$--30~M$_{\odot}$, this efficiency will be close to 100\% \citep{mt08}.
It is therefore difficult to avoid the conclusion that the final mass of the Pop.\ III
star will be very large.

Over the past few years, however, a new generation of numerical simulations
have begun to suggest that the true picture of Pop.\ III star formation may be
rather more complicated than the simple scenario outlined above. Work by
a number of groups (\citealt{cgk08,mach08,mach09,tao09,sgb10}) has found 
evidence for fragmentation in high density primordial gas, with predictions of the likelihood
that fragmentation occurs ranging from roughly 20\% \citep{tao09} to essentially
100\% \citep{cgk08,sgb10}. If fragmentation, and the formation of binary or multiple
systems, is indeed a common outcome of Population III star formation, then this
has profound implications for the final masses of the Pop.\ III stars, their production of 
ionizing photons and metals, the rate of high redshift gamma ray bursts, and many 
other issues.

It is therefore important to better understand the physical basis of
fragmentation in these systems. Although numerical simulations starting from
cosmological initial conditions and following collapse all the way to protostellar 
densities will give us the most accurate picture of how the first stars formed,
these simulations are computationally costly (making it difficult to explore a
wide parameter space), and are also difficult to interpret, since many physical
effects combine to yield the observed behaviour. Simulations that start from simpler 
initial conditions and that focus on exploring the importance of a single free 
parameter therefore play an important role, which is complementary to that of 
fully cosmological simulations. A good example is the recent work by Machida 
and collaborators \citep{mach08,mach09} which examined the influence
of the initial rotational energy of the gas, and found that zero metallicity
clouds with sufficient initial rotational energy could fragment into tight
binaries. In this contribution, we briefly describe the results of a similar
study that we have recently performed that looked at the effects of the
initial turbulent energy of the gas \citep{clark10}. 

\section{Simulations}
\subsection{Numerical method}
\label{sec:gadget2}
We modeled the evolution of the  primordial gas using a modified version
of the Gadget 2 smoothed particle hydrodynamics code \citep{springel05}.
Our modifications include the addition of a sink particle implementation
\citep{bbp95,jap05}, allowing us to follow the evolution of the gas beyond 
the point at which the first protostar forms, and a detailed treatment of 
primordial cooling and chemistry \citep{ga08,clark10}. Full details can be
found in \citet{clark10}.

\subsection{Initial conditions}
\label{sec:ics}
We took initial configurations that were unstable
Bonnor-Ebert spheres, into which we injected a subsonic turbulent velocity field. In our
study, we considered two different sets of initial conditions: one
corresponding to the Pop.\ III.1 scenario described above, and another corresponding to the
Pop.\ III.2 scenario, in which efficient HD cooling leads to a lower $T_{\rm ch}$ and
higher $n_{\rm ch}$ for the primordial prestellar core. However, in this contribution we
will discuss only the results of the Pop.\ III.1 simulations. In these, we set the initial gas temperature 
to be 300~K, and took  an initial central number density of $10^{5}$ cm$^{-3}$. The mass of the
Bonnor-Ebert sphere was taken to be $1000 \: {\rm M_{\odot}}$, corresponding to roughly three
Jeans masses. This mass was represented by 2 million SPH particles, giving us a particle mass
of $5 \times 10^{-4}  \: {\rm M_{\odot}}$, and a mass resolution of  $5 \times 10^{-2} \: {\rm M_{\odot}}$. Within the BE spheres, we imposed a turbulent velocity field that had a power spectrum $P(k) \propto k^{-4}$. The three-dimensional rms velocity in the turbulent field, $\Delta v_{\rm turb}$, was scaled to some fraction of the initial sound speed \cs. For the simulations presented here, we used four different rms velocities: 0.1, 0.2, 0.4 and 0.8~$c_{\rm s}$. Sink particles were created once the number density of the gas reached  $10^{13}$ cm$^{-3}$, using the standard \citet{bbp95} algorithm, with a sink accretion radius of 20~AU.

\section{Key results}
We find that the clouds with $\Delta v_{\rm turb} \geq 0.2 c_{\rm s}$  all fragment into small clusters of sink particles: the runs with  $\Delta v_{\rm turb} = 0.2 c_{\rm s}$, $0.4 c_{\rm s}$ and $0.8 c_{\rm s}$ have formed 5, 31, and 15 sink particles, respectively, at the point at which 10\% of the cloud mass has
been accreted. This clearly demonstrates that turbulent subsonic motions are able to promote fragmentation in primordial gas clouds. Only in the case with $\Delta v_{\rm turb} = 0.1 c_{\rm s}$ does the cloud form just a single sink particle. In this case, an extended disk builds up around the star,  but otherwise the gas contains little structure. 
 
Interestingly, there is no clear trend linking the number of fragments that form and the initial turbulent energy: the 0.4 \cs run fragments more than the 0.8 \cs run, despite containing only one quarter of the initial turbulent energy. The effects at play here are somewhat complex. First, the turbulent velocity field contained in the collapsing region will differ with each value of $\Delta v_{\rm turb}/c_{\rm s}$, since the different strengths of flow will push the gas around to different degrees. In addition, the nature of the turbulence that survives in the collapsing core will also affect the fragmentation.  The cloud with $\Delta v_{\rm turb} = 0.4 c_{\rm s}$ -- the most successful in terms of fragmentation -- forms a large disk-like structure, which appears to aid the fragmentation of the infalling envelope. 
 Much of the cloud's ability to fragment therefore seems to depend on the
amount of angular momentum that becomes locked up in the collapsing region, and hence on the
detailed properties of the turbulent velocity field. Further, the extra delay in the collapse caused by the increased turbulent support also gives the cloud more time to wash out anisotropies. The ability of the cloud to fragment is a competition between these conflicting processes. For the randomly generated velocity field that we used in this study, the ability to fragment is better for $\Delta v_{\rm turb}/c_{\rm s} = 0.4$, than $\Delta v_{\rm turb}/c_{\rm s} = 0.8$, but we note that this may not always be the case. 

\begin{figure}[t]
    		\includegraphics[height=2.5in]{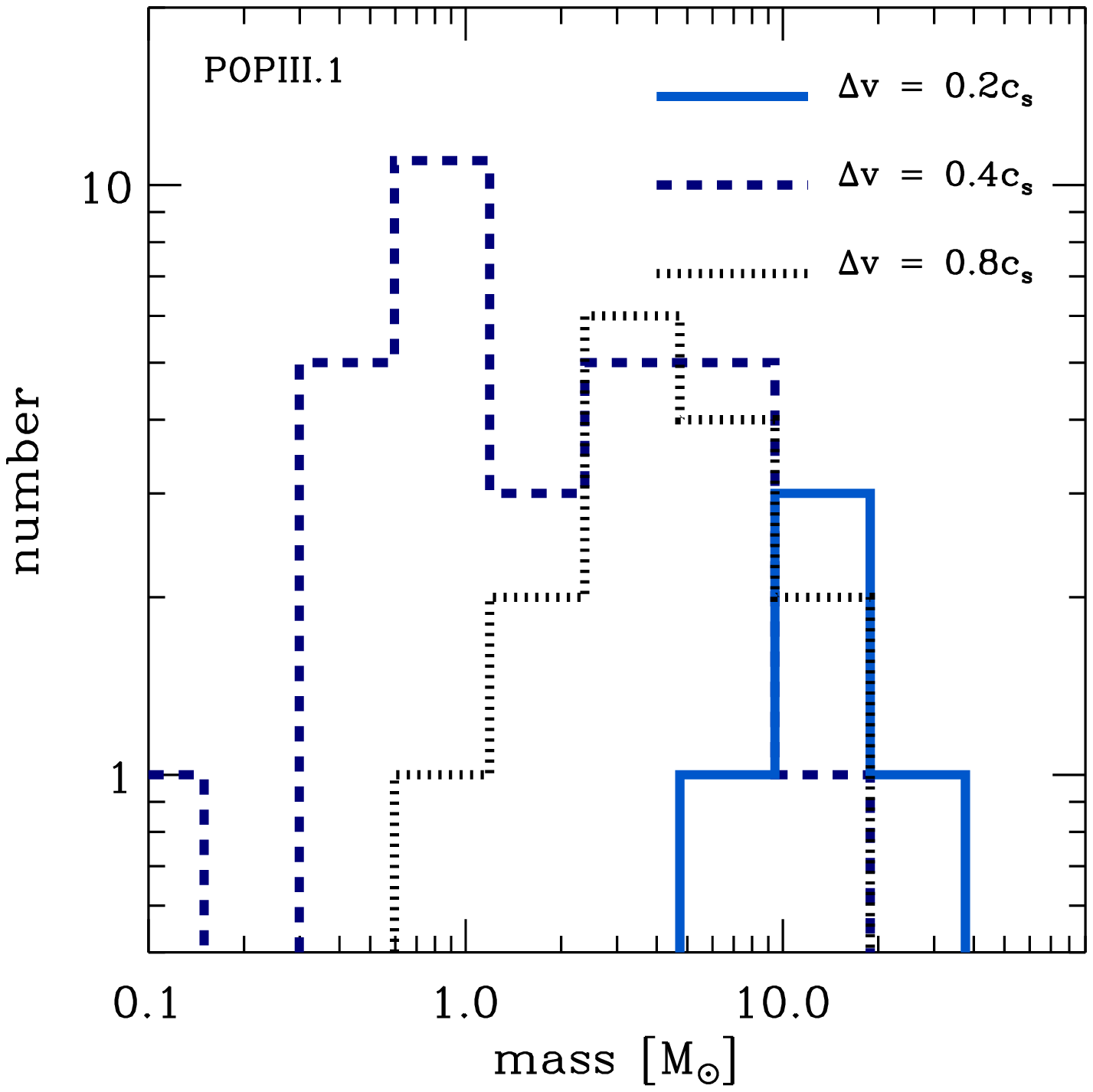}
		\includegraphics[height=2.5in]{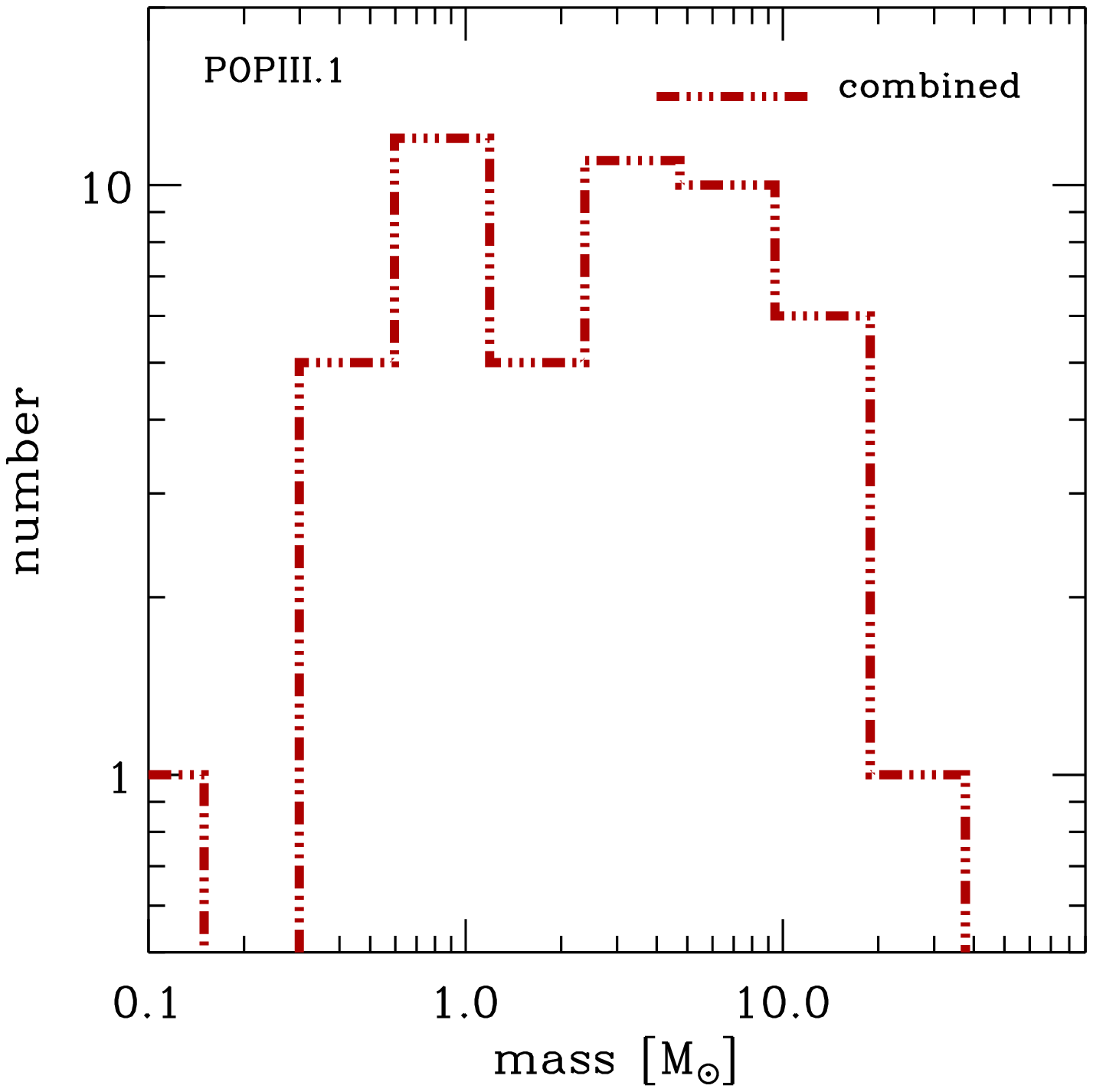}
\caption{\label{fig:sinkmf} The top panel shows the mass functions from the simulations in which fragmentation occurs. In all cases the mass function is plotted at the point at which the total
mass of gas converted to sink particles is 100\solmasp. Note that as accretion is ongoing, and the system is still young ($t \sim 1000$~yr), these will often not be the final masses of the sinks. The mass functions in the individual simulations differ substantially, although the combined mass function, shown in the bottom panel, exhibits a broad and flat distribution between masses of 0.4 and 20 \solmasp.}
\end{figure}

The mass functions of the sink particles from the simulations that undergo fragmentation are shown in Fig.~\ref{fig:sinkmf}. For clarity, we have omitted the single 100 \solmas sink particle that forms in the 0.1 \cs cloud. For the 0.2 \cs and 0.8 \cs  clouds, we see that the sink masses cluster around some central value -- roughly 12 \solmas and 4 \solmas respectively -- while the 0.4 \cs cloud has a mass function that is skewed to lower masses, with a peak at around 1 \solmasp,  a sharp fall-off below this, and a broad distribution towards higher masses, extending up to around 13 \solmasp. As all of the sinks form with
essentially the same initial mass, the spread of sink masses at the end of the simulation reflects the fact
that some sinks are more effective than others at accreting gas from the available reservoir within the
cloud. This behaviour is very familiar from studies of present-day star formation \citep[see e.g.][]{bvb04,sk04}, and is a result of the the high sensitivity of the mass
accretion rate to the sink mass and the relative velocity between sink and gas \citep{bonnell01a}:
$\dot{m_{*}} \propto \rho \, m_*^2 / v_{\rm rel}^3$, where $m_{*}$ is the mass of the sink particle, $\rho$ is the gas density, and $v_{\rm rel}$ is the velocity of the sink particle relative to the gas. As the number of  sinks increases, more and more 
get `kicked' by close dynamical interactions, forcing them onto more distant orbits, and increasing
$v_{\rm rel}$. These sinks therefore accrete very little gas following the
dynamical kick.  The few sinks that are not strongly kicked therefore accrete the lion's share of the available gas, and form the high-mass tail of the resulting mass function.  This process is termed `competitive accretion', \citep{bonnell01a} and normally leads to the type of distribution of masses seen in our 0.4\cs run. 

Our simulations therefore demonstrate that if significant levels of subsonic turbulence are present in the primordial gas found within the first dark matter halos, then Pop.\ III stars forming in these halos are more likely to be born in small stellar groups than as single stars. Moreover, if Pop.\ III stars do indeed form in this way, then we would expect the resulting initial mass function to be broad, thanks to the influence of competitive accretion within the star cluster. Tests of these expectations will require large,
high-resolution studies of the formation of the first  dark matter minihalos that are capable of resolving the turbulent flows in the gas at the moment when the baryons become self-gravitating, and that use
sink particles (or some comparable technique) to allow the evolution of the self-gravitating system to
be followed over multiple dynamical times. Such simulations would help determine which, if any, of the initial conditions presented in our study are actually realized in nature.

\bibliographystyle{apj}

\end{document}